\documentclass[12pt]{article}
\usepackage{amsmath,amssymb,bm,bbm,mathrsfs,amscd,natbib}
\usepackage{calc}
\usepackage{color}
\usepackage[margin=2.5cm]{geometry}
\usepackage{amsthm,placeins}
\usepackage{xurl}
\usepackage{hyperref,cite}
\usepackage{graphicx}
\usepackage{mathptmx}       
\usepackage{graphicx}        
\usepackage{multicol}        
\usepackage[bottom]{footmisc}
\usepackage{subfig}
\usepackage{calc,xcolor}
\usepackage{color}
\renewcommand{\eqref}[1]{Eq.~(\ref{#1})}  
\usepackage{hyperref}  

\begin{document}
\date{}
\title{Data-informed modeling of the formation, persistence, and evolution of social norms and conventions}

\author{\centering\large Mengbin Ye$^1$ and Lorenzo Zino$^2$}
\maketitle

\date{\vspace{.5cm}

\normalsize 
\noindent $^1$Centre for Optimisation and Decision Science, Curtin University, Bentley 6149 WA, Australia\\
$^{2}$Department of Electronics and Telecommunications, Politecnico di
Torino, 10129 Torino, Italy\\\vspace{.2cm}

\noindent Correspondence should be addressed to: \url{mengbin.ye@curtin.edu.au}}

\begin{abstract}Social norms and conventions are commonly accepted and adopted behaviors and practices within a social group that guide interactions ---e.g., how to spell a word or how to greet people--- and are central to a group's culture and identity. Understanding the key mechanisms that govern the formation, persistence, and evolution of social norms and conventions in social communities is a problem of paramount importance for a broad range of real-world applications, spanning from preparedness for future emergencies to promotion of sustainable practices. In the past decades, mathematical modeling has emerged as a powerful tool to reproduce and study the complex dynamics of norm and convention change, gaining insights into their mechanisms, and ultimately deriving tools to predict their evolution. The first goal of this chapter is to introduce some of the main mathematical approaches for modeling social norms and conventions, including population models and agent-based models relying on the theories of dynamical systems, evolutionary dynamics, and game theory. The second goal of the chapter is to illustrate how quantitative observations and empirical data can be incorporated into these mathematical models in a systematic manner, establishing a data-based approach to mathematical modeling of formation, persistence, and evolution of social norms and conventions. Finally,  current challenges and future opportunities in this growing field of research are discussed.\end{abstract}

\subsection*{Preprint Notice}
This is an author's (preprint) version of a book chapter that is part of the Handbook of Visual, Experimental and Computational Mathematics --- Bridges through Data: \url{https://doi.org/10.1007/978-3-030-93954-0}.
\newpage

\section{Introduction}\label{sec:intro}

{Mathematical modeling} has increasingly emerged as a powerful tool to describe, study, and predict complex social dynamics. Such models are applied to a broad range of applications; several examples are mentioned here to provide an impression, while the interested reader may consider review papers for broader coverage~\citep{Castellano2009,Jusup2022}. Different application contexts include, for instance, studying social influence and opinion formation~\citep{french1956_socialpower,degroot1974OpinionDynamics,anderson2019IJAC,Noorazar2020}, understanding the diffusion of social innovations~\citep{Bass1969,montanari2010spread_innovation,young2011dynamics,zino2022nexus} and the spread of misinformation~\citep{DelVicario2016,Franceschi2022}, anticipating crimes and violence~\citep{DOrsogna2015,Succar2024}, unveiling complex political processes~\citep{Leonard2021,Fontan2021}, and predicting the behavioral response of people to an epidemic outbreak~\citep{Cinelli2020,Ye2021pre}.

The use of mathematical models to study social phenomena can be traced back to pioneering efforts developed in the 1940s, which were not met with ready acceptance by social scientists at large. For a survey of such pioneering works, see~\citet{Rapoport1963,Abelson1967}. Mathematical models started to become widely used and accepted in the social science community from the end of the 1960s, thanks to an array of seminal works including the diffusion model proposed by Frank Bass~\citep{Bass1969}, the opinion dynamics model proposed by Morris H. DeGroot~\citep{degroot1974OpinionDynamics}, and the linear threshold model due to Mark Granovetter~\citep{granovetter1978threshold}. These works paved the way for a flourishing growth of the field of mathematical modeling of social systems over the past several decades, witnessing the development of more refined mathematical models tailored to reproduce specific phenomena of interest~\citep{Edling2002,Castellano2009,Jusup2022}. In particular, a crucial advancement can be pinpointed to the design and refinement of agent-based models~\citep{Bonabeau2002abm}, in which a population of  heterogeneous agents interact following simple agent-based rules  on complex network structures~\citep{VegaRedondo2007,Easley2010}. This modeling paradigm allows one to encapsulate within the mathematical model key features of human behavior, such as the tendency to be influenced by their peers in a non-additive manner and the non-homogeneous and often time-varying pattern of social interactions~\citet{Holme2012,guilbeault2018complex}. 

 In order to move beyond using such models to explain observed phenomena and towards direct application, a crucial challenge is the integration of empirical observation and data into the design and validation of the models, and in their calibration to real-world scenarios. In the past, models were typically developed by first proposing mechanisms that were grounded or inspired by established social psychology theories and general observations drawn from empirical evidence. Then, the models were validated in controlled experiments and fitted to field data. This is the case, for instance, of classical opinion dynamics models, which are based on the mathematical mechanism of linear averaging. The seminal DeGroot model of opinion formation~\citep{degroot1974OpinionDynamics} was built on the social psychological theories of social influence and social power~\citep{french1956_socialpower}. It has been validated experimentally subsequently~\citep{Becker2017,Chandrasekhar2020} and fitted to online social media data~\citep{Kozitsin2021}. Similarly, one of its most successful extensions, the Friedkin--Johnsen model, was proposed based on the social psychological theory that people are unwilling to depart from existing prejudices~\citep{friedkin1990social}, and has since  then been validated experimentally~\citep{Friedkin2017} and fitted to the discussion dynamics of the Paris Climate Accords~\citep{Bernardo2021}.  
 
Recently, the unprecedented availability of large datasets (e.g. online social media data) and the tremendous advancements in  data science have dramatically changed the picture around modeling social systems, leading to the birth of the novel emergent field of {computational social science}~\citep{Lazer2009}. Besides the development of model-free data-driven approaches, even in the context of model-based approaches such a paradigm shift has increased the  centrality of data integration into mathematical models. In this context, data are not just used a posteriori to calibrate existing models as noted above, but also to inform the design and refinement of new and improved models. This ultimately yields a data-based approach to mathematical modeling of social systems. 
 
The previous paragraphs have illustrated how the field of mathematical modeling of social systems is extremely broad and continuously evolving, with the continuous development of new models, methodologies, and applications. This chapter focuses on the specific application to the formation, persistence, and evolution of social norms and conventions. In other words, the main focus of the rest of this chapter is on mathematical models tailored to capture and reproduce the collective adoption, maintenance, and replacement of a behavior, idea, or action by a social community~\citep{peytonyoung2015social_norms}. Shedding light onto this phenomenon and developing mathematical models that are able to predict it and, ultimately, control said phenomenon have important societal implications in the context of social change towards a sustainable economy~\citep{Selin2021,Hoffmann2024} and preparedness to future emergencies and disasters~\citep{Bavel2020}. The aim of this chapter is to introduce the reader to data-driven mathematical modeling of social norms and conventions, and as such, the chapter will elaborate on several illustrative examples rather than provide a comprehensive account of the literature.

The rest of the chapter is organized as follows. 
First, the problem of modeling social norms and conventions is presented, with a brief discussion of these concepts from a psychological and sociological perspective, illustrated by means of real-world examples and highlighting their key features. 
Second, the main mathematical approaches used to model these social phenomena are discussed. The chapter starts presenting population models, which are the first and simplest class of models, and highlighting their inherent limitations in capturing key features of social systems, such as interpersonal interactions, social influence, and heterogeneity. Then, different classes of agent-based models that have been developed to overcome the limitations of population models are presented:  cascading models, evolutionary dynamic models, and game-theoretic decision-making models. 
Third, the use and integration of real-world data to design, inform, validate, and calibrate these mathematical models is extensively discussed, illustrated by different examples for different classes of agent-based models. 
Fourth, the chapter is concluded by providing a summary of the main advances in the modeling of formation and evolution of social norms and conventions and on the integration of real-word data, and discussing current and future challenges and opportunities in this promising research field.

\section{Background on norms and conventions}\label{sec:background_norms}

In order to develop appropriate mathematical models of social norms and conventions, one first needs to understand what norms and conventions are from a theoretical psychological and sociological perspective. For this reason, this section is devoted to a brief discussion of relevant concepts,  with explanations by means of classical examples from the related literature concerning the formation, persistence, and evolution of norms and conventions.

{Social norms} and {social conventions} are fundamental aspects of our societies and their cultures, and help to provide a framework for interactions between people~\citep{Lewis2002,bicchieri2014norms,Gelfand2016}. Despite  conceptual and functional differences between norms and conventions which are theoretically debated upon in the social psychology literature~\citep{Southwood2011}, they have some important common traits. Thus, from a modeling perspective, one can consider norms and conventions together. Throughout this chapter, the term ``norm'' and ``convention'' will be used interchangeably. A common feature of both is that their value mostly depends on their widespread adoption and acceptance~\citep{Bicchieri2005,bicchieri2014norms,Lewis2002}. For instance, when walking on a crowded sidewalk, it matters little whether one walks on the left-side or right-side of the sidewalk, provided that others do likewise~\citep{Lewis2002}. When greeting others, there is not much difference between the use of handshakes versus bowing, but it would be embarrassing to not coordinate on the same gesture~\citep{Marmor2009}. When collaboratively writing a document in English, especially for instance a scientific manuscript featuring authors from different countries, one could use British or American spelling, as far as the document is consistent and all the authors use the same spelling~\citep{lieberman2007quantifying}. These are all examples of social conventions, where individuals benefit from coordinating with peers to adopt the same behavior among several equivalent alternatives. Norms share many similarities, but one key difference is that the the normative behavior (whatever is accepted as the standard in the group) is often enforced by potential or actual punishment for deviants. For instance, it is now a generally accepted norm within warfare that medical personnel and wounded soldiers are considered noncombatants~\citep{finnemore1998international}, and is now part of the 1949 Geneva Convention. The processes of formation, persistence, and evolution of social norms are fundamental for the well functioning of our societies~\citep{Lewis2002,Marmor2009}. The understanding of these key societal processes is critical for public authorities who, for many different reasons, may be interested in changing the  conventions or social norms currently adopted by a population in a systematic manner, e.g., to promote a persistent change towards more sustainable behaviors to face the ongoing climate change crisis~\citep{Selin2021} or being able to promptly adapt conventions and norms to respond to societal emergencies, as happened during the COVID-19 health crisis~\citep{Bavel2020}.

\subsection{Real-world examples}\label{ssec:norm_examples}

Here, some examples of real-world social norms  and conventions are presented and discussed. These examples illustrate some of the key features of norms and conventions and show how they form, persist in time, and evolve. Readers interested in further examples can enjoy the excellent survey by \citet{peytonyoung2015social_norms}.

\subsubsection{Footbinding in rural China} {Footbinding} practice among Chinese women is a classical example of the evolution of a social norm, extensively studied in the sociology literature~\citep{mackie1996footbinding,brown2018economic}. This practice involved the painful application of a binding cloth over a girl's feet (often from a young age) for a long period, in order to change their shape and reduce their size. Those that did not undergo footbinding were seen as less desirable in terms of marriage, illustrating the disadvantages in deviating from the norm. It is recorded that footbinding was the normative practice in rural China for many centuries, during the Ming Dynasty (1368-1644), and that efforts to abolish such a practice started in 1665 by the Manchu government~\citep{mackie1996footbinding}. However, footbinding persisted as a social norm for several centuries: at the beginning of the 20th century, the vast majority of women in Northern rural counties underwent footbinding~\citep{mackie1996footbinding}. It was only in the 1910s that women in Northern and Central China started abandoning footbinding, with a complete change of the social norm occurring over a 30-year span. This famous example illustrates some key aspects of social norms, such as the value associated with their widespread adoption, their persistence, and their sudden change.

\subsubsection{Spelling of ``s\'olo" vs ``solo" in Spanish}

\begin{figure}[t!]
    \centering
\subfloat[]{\includegraphics[height=5cm]{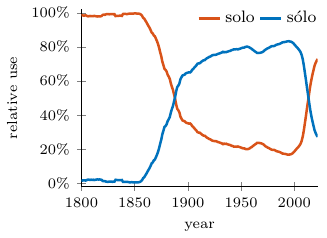} \label{fig:solo}}\qquad\subfloat[]{\includegraphics[height=5cm]{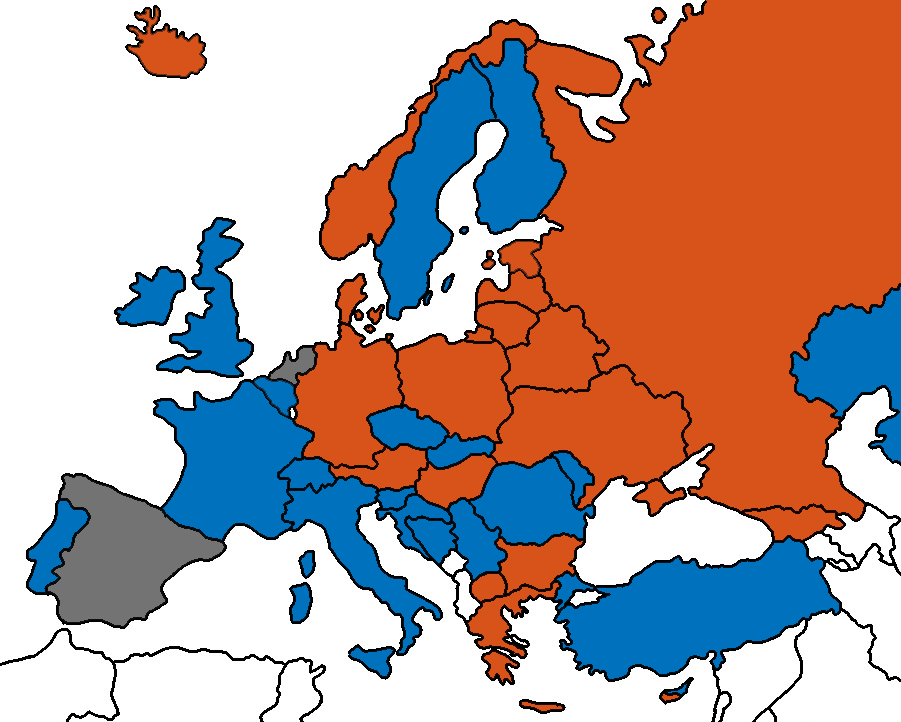} \label{fig:map}}  \caption{Examples of formation, persistence, and evolution of social norms and conventions. In panel (a), relative use of ``s\'olo" (blue) and ``solo" (orange) in written Spanish from  Google Ngram data~\citep{Michel176}. In panel (b), map of Europe illustrating the hand used for wedding ring in different countries: blue for vast majority of left hand, orange for vast majority of right hand, gray for regional/cultural differences within the country. }
    \label{fig:enter-label}
\end{figure}

Other relevant examples of conventions come from the field of linguistic. In fact, the {spelling} of a specific word or even the meaning of a word do not have intrinsic value. Rather, their value is clearly associated with agreement on a specific spelling or interpretation within a community. An example often used in this literature is the spelling of the adverb ``only" in Spanish. In fact, two spellings (``s\'olo" and ``solo") have been used over the past two centuries~\citep{Amato2018}. At the beginning of the 19th century, the only spelling used was ``solo." Then, according to Google Ngram data~\citep{Michel176}, the use of ``s\'olo" quickly became dominant at the end of the 19th century and remained preferred for almost a century, until the 2010s, when a swift change of convention reverted to the spelling ``solo", potentially as it is more convenient on US keyboards. This evolution is illustrated in Fig.~\ref{fig:solo}.

\subsubsection{Hand on which to wear a wedding ring}

Another interesting example of convention change is the use of the {wedding ring}, which can be traced back to the Roman Empire or even earlier, to the Ancient Egyptians. While it is common in all Western countries to wear a ring (typically made of gold) on the fourth finger (called ``ring finger" for this reason) to indicate that its wearer is married, there is not global agreement on which hand to wear it~\citep{monger2004}. These differences are illustrated in Fig.~\ref{fig:map}. Interestingly, the figure shows a pattern with high local conformity, but global diversity. In fact, in Europe, the vast majority of the population within a given country wears the wedding ring on the same hand. However, there is diversity between different countries, due to different cultural factors. There are two exceptions where there is no agreement within the same country, but even in these cases, there is strong local conformity. In the Netherlands, Protestants wear their wedding ring on the right hand, while Catholics on the left; in Spain, wedding rings are typically worn on the right, except for Catalonia and adjacent regions.

\subsection{Key features of norms and conventions}\label{ssec:norm_features}

The examples discussed in the previous section have highlighted some key universal features of social norms and conventions that one needs to keep in mind in order to develop mathematical modeling frameworks that can reproduce and predict their formation, persistence, and evolution. 

First, as already discussed in the above, a key feature of social norms and conventions is the fact that their value depends on their widespread adoption and acceptance. Thus, norms and conventions are often enforced by an individual's tendency and desire to act in {conformity} with societal expectations~\citep{marques1994,VegaRedondo2007}. In fact, there is no clear advantage in using the spelling ``s\'olo" instead of ``solo", besides conforming with others. Moreover, it is important to notice that some social norms can be characterized by  \textit{global diversity but strong local conformity}, as in the example of the use of the wedding ring, where the hand on which the ring is worn can be different in different communities/countries, but are locally consistent within the same community.

Second, in many of the examples discussed in the above, it is possible to observe the {persistence} of a social norm or convention over a long period. In fact, footbinding persisted as the norm in rural China for centuries, even when the government tried to ban it~\citep{mackie1996footbinding}. However, a norm or convention cannot persist indefinitely for all time, for obvious reasons. From a modeling point of view, this can be associated with the mathematical concept of meta-stability. In fact, many real-world examples feature a replacement of the status quo norm or convention by an alternative (e.g., footbinding replaced by non-footbinding, or the spelling ``solo" replaced by ``s\'olo"). Interestingly, replacement often occurs over a time window that is substantially shorter when compared to the phase in which the status quo persisted, so that the fraction of adopters of the alternative convention  typically follows an {S-shaped curve}, following the terminology of the existing literature~\citep{rogers2003diffusion} (see, e.g., Fig.~\ref{fig:solo}). Finally, it is important to also notice that this replacement itself will not remain as the new permanent convention, but rather,  it becomes the status quo for a period before also likely being replaced by another alternative (or even by the original one, as in the spelling example), in a continuous process of social innovation that is key for the functioning and progression of our societies.

The third important feature, which is really at the core of this chapter, is that the collective adoption and evolution of social norms and conventions is typically the emergent behavior of a population, whereby individuals engage in repeated interactions over time, through which one individual can observe the behavior of others (e.g., by exchanging messages or e-mails, one becomes aware of how others spell a word; or by meeting, one can observe how others greet) and is influenced by such observations~\citep{peytonyoung2015social_norms}. The effect of these interactions is often complex and non-linear. It is known from many empirical studies that the pattern of social interactions is inherently complex, characterized by spatial and temporal heterogeneity, and can be conveniently captured by {complex network} structures~\citep{Boccaletti2006}. This observation explains the success of agent-based models within this field. In fact, in order to capture this important feature of social norms, one needs to have a mathematical description of the dynamics at the granularity of an individual. Building on this individual-level description, a mathematical model that describes how the individual behavior is affected by the interpersonal interactions which occur on a complex network is derived. In agent-based models, these interactions result in an emergent behavior at the population level that can be highly organized, counter-intuitive, and both robust and fragile to system shocks.

\section{Mathematical modeling of norms and conventions}\label{sec:modeling}

This section illustrates and discusses some of the main mathematical approaches used to model the formation and evolution of social norms and conventions. The development of these approaches began from the empirical observations of real-world instances of the formation and evolution of norms and conventions. From these observations, the mathematical models that are developed aim to encapsulate some of the key features of social norms that have been highlighted and extensively discussed in the previous section of this chapter. The models considered aim to balance (arriving at different points on the spectrum) between being sufficiently complex to capture the desired phenomena and sufficiently simple so as to be tractable (both analytically and numerically). The models also frame the problem in a similar manner: initially, there is widespread adoption within the population of one norm or convention, termed the \textit{status quo}, and a new norm or convention is introduced into the population, termed the \textit{innovation} or \textit{alternative} (here, innovation refers to the fact that it is novel to the population, and not necessarily an entirely new norm). 

This section is divided into two main parts. The first part  focuses on discussing population models, where the formation and evolution of social conventions is described at the granularity of the entire population. In other words, the proposed models keep track of the temporal evolution of the population fraction adopting the innovation. The first and most famous model belonging to this class is the Bass model~\citep{Bass1969}, which is extensively presented and discussed in the sequel. Population models, however, have inherent limitations as they struggle at capturing the complex and heterogeneous pattern of interpersonal interactions and social influence that drive social change, as explained in detail below. For this reason, the second and larger part of this section is devoted to agent-based models. These reproduce the phenomenon of formation and evolution of social conventions as the emergent behavior of a population of individuals (agents), whose dynamics are described at the granularity of the individual~\citep{Bonabeau2002abm}. In particular, this section presents and discusses four classes of agent-based models that have been extensively used in the context of social norms and conventions: cascading models, evolutionary dynamic models, and game-theoretic decision-making models.

In many fields, one can often rely on established laws or principles within said field to derive the equations to describe the dynamics of the system of interest. For instance, the dynamical equations for electromagnetic circuits can be derived from Maxwell's equations, while the equations of motions of mechanical systems can be derived from Newton's laws of motion. Even in mathematical epidemiology, models of infectious disease spread are typically derived following mass conservation principles. In contrast, while scientists have established theories and observations (qualitative and quantitative) to describe human behavior, there are no laws or principles in the same sense as in electromagnetism or mechanics. Thus, many models of social dynamics involve formulating terms within the equations to conceptually represent social and behavioral mechanisms of relevance. For the models presented below, the details of the derivations will be provided for some of them to give context. For others, explicit derivations are omitted due to their complexity (giving instead references to articles or books that contain a complete mathematical derivation), but  descriptions of the intuition underpinning the specific equations are provided.

\subsection{Population models}\label{ssec:popn_models}

{Population models} are the first class of mathematical models developed to study the adoption and evolution of conventions and norms. In this model class, the focus is in studying and reproducing the social phenomenon of interest with a mathematical description of the system at the population level. In other words, by defining a variable $z(t)\in[0,1]$ that measures the fraction of adopters of a specific norm or convention at time $t$. For consistency, in this chapter all models will be presented as discrete-time dynamics (i.e, $t\in\mathbb N_{\geq 0}$). However, one should keep in mind that there are continuous-time formulations for most of the models presented in the following. The population model establishes a mathematical rule that determines how $z(t)$ evolves in time, ultimately yielding a dynamical system that is analytically tractable due to its low dimensionality. Different mathematical approaches can used to define the evolution of the variable $z(t)$, including ordinary differential equations, recursive equations, and stochastic processes. 

One of the first and most successful models developed within the framework of population model is the famous {Bass diffusion model} proposed by Frank Bass in~\citet{Bass1969}. This model is grounded in the marketing and management science theory of innovation diffusion developed by Everett M. Rogers from the early 1960s~\citep{rogers2003diffusion}, and was initially proposed within a marketing framework to study how new products are adopted by a population. Since then, it has been generalized and applied to many scenarios involving a population adopting an innovation, including a new social norm or convention; see, e.g.,~\citet{
mahajan1990bass_survey,Horvat2020}. 

\begin{figure}[t]
    \centering
\subfloat[]{\includegraphics[height=6cm]{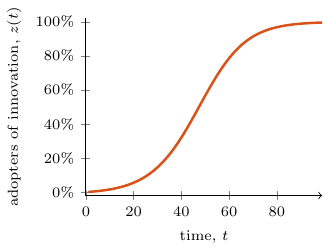}}\quad\subfloat[]{\includegraphics[height=6cm]{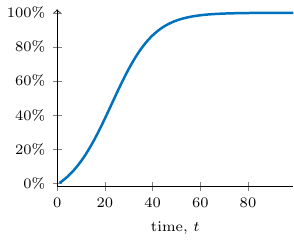}}
\caption{Two exemplary trajectories of the adoption curve obtained with the discrete-time Bass model in \eqref{eq:bass} with (a) $p=0.001$ and $q=0.01$; and (b) $p=0.001$ and $q=0.1$. Both plots reproduce an S-shaped curve, with different initial slope and velocity to reach the inflection point. }
    \label{fig:bass}
\end{figure}

This model consists of a recursive equation that describes how $z(t)$ evolves in time. Specifically, the temporal evolution of the fraction of adopters is determined by the following nonlinear autonomous recursive equation: \begin{equation}\label{eq:bass}
    z(t+1)=z(t)+p\big(1-z(t)\big)+qz(t)\big(1-z(t)\big),
\end{equation}
where $p,q\in\mathbb R$ are two constant parameters (typically non-negative in healthy markets). In plain words, \eqref{eq:bass} establishes that the fraction of adopters of the innovation grows according to two distinct mechanisms. Besides current adopters captured by the first term, the second term on the right hand side of \eqref{eq:bass}, $p(1-z(t))$, accounts for new adoptions via innovation: a constant fraction of those who are not yet adopting the innovation will decide to adopt it. Hence, such a term is proportional to the fraction of non-adopters (i.e, $1-z(t)$), and multiplied by the parameter $p$ that represents the strength of the innovation process. The third term, $qz(t)(1-z(t))$, instead, accounts for imitation: non-adopters who interact with adopters are then convinced to adopt at a given rate $q$. For this reason, the second term is non-linear, given by the product between adopters ($z(t)$) and non-adopters ($1-z(t)$). Observe that the solution of the recursive equation in \eqref{eq:bass} is a logistic equation, which is characterized by a S-shaped curve, as illustrated in Fig.~\ref{fig:bass}. It is worth noticing that the original formulation of the model was in continuous-time as a differential equation, while the formulation in \eqref{eq:bass} is instead based on~\citet{Satoh2001}.

In his seminal paper, Frank Bass fitted his model by calibrating the parameters $p$ and $q$ via statistical regression, using real-world marketing data on sales of different innovative products in the decades before the publication of \citet{Bass1969}, such as home freezers and black-and-white televisions. Through such a fitting, he obtained respectable approximations of the empirically observed S-shaped adoption curves~\citep{rogers2003diffusion}.  Subsequent studies have extended the validation using different datasets from different application fields~\citep{Sultan1990}.

Despite its ability to reproduce real-world adoption curves, the Bass model is subject to some critical limitations. The fact that it provides a description of the phenomenon only at the population level prevents the model from encapsulating some of the features mentioned above, including the key role of interpersonal interactions, social influence, and heterogeneity across the population. These shortcomings limit the possibility to adopt the Bass model (and, more in general, population models) to study how the complex network of social interactions and the inherent complexity of the behavioral factors influencing individual-level decision-making mechanisms can affect the adoption curve. In particular, the shortcomings become especially noticeable when, taking on the perspective of a policymaker or authority, one wishes to apply individual-level interventions (e.g. monetary incentives) to facilitate widespread adoption of the innovation. Since the granularity is at the population level, such interventions cannot be easily incorporated. These limitations have thus led to the development of a new class of mathematical models that describe the social dynamics at the granularity of the  individual.

\subsection{Agent-based models}\label{ssec:abm}

In the past few decades, {agent-based models} have emerged as a powerful approach to go beyond the limitations of population models, being able to reproduce the adoption process at the level of an individual. Such an approach allows one to study how individual-level mechanisms affect the population-level behavior, providing a novel array of tools to study the role of individual social and psychological factors in the formation and evolution of social norms and conventions. Importantly, such models are also better suited to studying the effects of intervention policies.

Agent-based models can differ in many aspects concerning the rationale and the mathematical implementation of the mechanisms that govern each individual's behavior. However, they all share key common traits. First, all agent-based models consider a population of $n$ agents (individuals), which are denoted by integer indices, i.e., the set $\mathcal V:=\{1,\dots,n\}$. Each individual $i\in\mathcal V$ is characterized by a state $x_i(t)$, which represents the choice of individual $i$ with respect to the norm or convention considered in the model. This chapter considers the simplest scenario, involving the diffusion of a single innovation (alternative norm or convention) in a population, and thus the model posits that $x_i(t)$ is a binary variable, defined as
\begin{equation}\label{eq:state}
    x_i(t)=\left\{\begin{array}{ll}1&\text{if $i$ adopts the innovation at time $t$,}\\
    0&\text{if $i$ does not adopt the innovation at time $t$.}
    \end{array}\right.
\end{equation}
Further nuances can be introduced, e.g. by allowing $x_i(t)$ to take a value from a discrete set (if multiple options are possible), or assuming it to be continuous. Given the formulation in \eqref{eq:state}, the adoption curve can be observed as an emergent behavior of the dynamics of the agents. In fact, the total fraction of adopters is equal to
\begin{equation}
    z(t)=\frac{1}{n}\sum_{i\in\mathcal V}x_i(t).\end{equation}

A second common feature of agent-based models is the fact that, as reflected in the real world, individuals interact with their peers on a {complex network} structure, exchanging information on their state and mutually influencing one another. Real-world social networks have been extensively investigated utilizing large datasets from online social networks~\citep{Fu2008} and face-to-face experiments with wearable proximity sensors~\citep{Cattuto2010}. These studies have allowed researchers to derive a clear picture of their structure and key characteristics, including heterogeneity~\citep{Newman2018}. Based on these empirical observations, agent-based models typically represent the interacting individuals as nodes of a network $\mathcal G=(\mathcal V,\mathcal E)$, where the edge set $\mathcal E\subseteq\mathcal V\times\mathcal V$ is such that an edge $(i,j)\in\mathcal E$ if and only if $i$ is influenced by $j$. If not differently stated, it is typically assumed that edges are bidirectional, i.e., the presence of edge $(i,j)$ means that $i$ is influenced by $j$ and $j$ is influenced by $i$. Hence, the set of neighbors of $i$, denoted by $\mathcal N_i(t):=\{j:(i,j)\in\mathcal E\}$ represents all agents who (directly) influence individual $i$. 

\begin{figure}[t]
    \centering
\includegraphics[width=\linewidth]{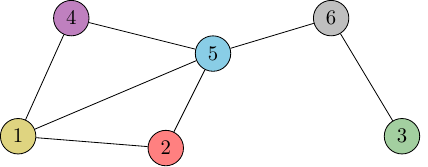}
\caption{Example of a network structure, with node set $\mathcal V=\{1,2,3,4,5,6\}$ and edge set $\mathcal E=\{(1,2),(1,4),(1,5),(2,5),(3,6),(4,5),(5,6)\}$. Note that node $5$ (in blue) has three neighbors ($\vert \mathcal N_5\vert=3$), namely $\mathcal N_5=\{2,4,6\}$. Hence, its state $x_5(t)$ evolves as a function of its own current state, and the states of its three neighbors. }
    \label{fig:network}
\end{figure}

Figure~\ref{fig:network} provides a simple illustration of a network structure. Different agent-based models can assume different network structures, which can be time-invariant, or evolve in time, even in a state-dependent manner~\citep{Holme2012}. For the sake of simplicity, this chapter will focus on models on time-invariant networks, while briefly mentioning some extensions to time-varying frameworks. It should also be noted that the edges connecting agents are abstract representations of {social influence}; edges exist if one agent can influence the state of another agent in some way, and edges can represent a range of interactions such as friendships, communication in online social media platforms, or agents observing each other in the physical world.

Finally, a third common trait of agent-based models is the presence of {agent dynamics}, which determines how the agent state evolves in time to become $x_i(t+1)$, as a function of its current state $x_i(t)$ and the state of neighboring nodes of the network, i.e., $x_j(t)$ with $j\in\mathcal N_i$. The main differences between different agent-based models typically lie in the agent dynamics, i.e., on how agents revise their state on the basis of some internal dynamics and due to social influence from their peers on the network.

A first, intuitive, modeling approach can be to embed the Bass model described in the previous section onto a network structure. This is done by assuming that the imitation mechanism present in the Bass model is regulated by a network, whereby an individual tends to imitate only those adopters among their neighbors. This approach has many similarities with epidemic models, whereby contagion (by a disease or by an innovation) is driven by spontaneous mechanisms (spontaneous recovery or adoption of the innovation) and pairwise interactions (transmission of the disease via contagion or imitation), and has been extensively adopted in the literature to derive social diffusion models~\citep{Rizzo2016,Bertotti2016,Fibich2016,Fagnani2017}. 

However, these imitation-driven models rely on the simplifying assumption that social dynamics evolve similarly to epidemics, according to a so-called {simple contagion} mechanism, whereby each interaction with an adopter (infected individual) yields a certain rate or probability of adopting (being infected), and the effects of each interaction are independent. On the contrary, it is known that many social processes follow a {complex contagion} mechanism~\citep{centola_book}. Here, multiple interactions with different neighbors are required to influence an individual, and interactions have, in general, a nonlinear impact on the dynamics. The rest of this section presents three important classes of agent-based models that differ in the agent dynamics but are all characterized by the presence of a complex contagion mechanism: namely, models based on threshold phenomena, evolutionary dynamics, and game theory.

\subsubsection{Linear threshold models}\label{sssec:cascade_model}

One of the first and most successful approaches proposed to capture a complex contagion mechanism within a mathematical modeling framework is the {linear threshold model}, proposed by Mark Granovetter~\citet{granovetter1978threshold}. This model was primarily proposed to reproduce the diffusion of an innovation  (e.g., a novel convention) within a network of individuals. Hence, individuals have a binary state representing whether they have adopted the novel convention or not, as in \eqref{eq:state}, and each individual is associated with a parameter $\theta_i\in[0,1]$ that represents the resistance of individual $i$ to change. In its simplest implementation, the linear threshold model is a discrete-time deterministic process. At each time-step, individual $i$ decides to adopt the innovation if and only if more than a fraction $\theta_i$ of their neighbors have already adopted the innovation, i.e., 
\begin{equation}\label{eq:ltm}
    x_i(t+1)=\left\{\begin{array}{ll}1&\text{if }\frac{1}{d_i}\displaystyle\sum_{j\in\mathcal N_i} x_j(t)\geq\theta_i,\\ 0&\text{otherwise},\end{array}\right.
\end{equation}
where $d_i:=|\mathcal N_i|$ is the number of neighbors (degree) of $i$ on the network. 

In this model, it is typically assumed that one or multiple seeds are present and act as innovators~\citep{rogers2003diffusion}, while all others agents are adopting the status quo. This can be easily modeled by setting the threshold of a seed node to $\theta_i=0$, which means that they adopt the innovation, regardless of what their neighbors do.
Then, the entire population is initialized to adopt the status quo ($x_i(t)=0$, for all $i\in\mathcal V$), and it can be easily proved that the state is monotonically increasing (i.e., if an individual switches to adopt the innovation, then they never revert to the status quo). In this setting, initial seeds may trigger a {cascade}, where at each time-step one or multiple agents may decide to adopt the innovation, in a recursive manner, resulting in the innovation spreading through the network. 

Researchers have studied the model in \eqref{eq:ltm} on different network structures, in order to understand how social interactions facilitate or hinder triggering diffusion cascades and affect the size of the cascade. For instance, one can refer to~\citet{Watts2002} for an extensive study on random networks and to~\citet{Rossi2019} for a study of the impact of the degree distribution of the network. Moreover, several extensions have been proposed including, e.g., time-varying thresholds~\citep{Arditti2024}. One popular optimization problem studied using the linear threshold model is known as the {influence maximization} problem~\citep{kempe2003_maxspread}. In its simplest form, a policymaker has the objective to maximize $\lim_{t\to\infty} z(t)$, i.e., the number of individuals who end up adopting the innovation, by selecting the network locations of a fixed number of seed nodes. Readers may refer to \citet{li2018influence} for an introductory survey.  

The linear threshold model and its extensions have been widely used to study innovation diffusion processes, largely thanks to their simplicity and intuitive implementation. However, they have some limitations. First, they typically do not allow individuals to drop the innovation and switch back to the status quo, which has been observed in many real-world examples. Second, in order to create a cascade leading to the diffusion of the innovation, \eqref{eq:ltm} typically requires heterogeneous thresholds $\theta_i$, which translates into a model with potentially many parameters, and thus difficult to be calibrated without the risk of overfitting. Third, despite being able to capture some features of complex contagion through the threshold mechanism, it is difficult to easily incorporate other behavioral mechanisms (besides imitation/social pressure) into the model, meaning key mechanisms that affect real-world decisions are omitted.

\subsubsection{Evolutionary dynamics models}\label{sssec:evolutionary_models}

Another approach to model social norms and conventions is by means of {evolutionary dynamics}. Evolutionary dynamics were originally proposed in the context of biological and ecological systems, to study the competition between different species and the evolution of genetic traits in the same species. In plain words, evolutionary dynamics consider a network of nodes (often representing geographical locations), which can be occupied by individuals of different species, who can interact with those in neighboring nodes and replace them according to a fitness mechanism. For a simple and intuitive formulation of evolutionary dynamics, one can refer to~\citet{Lieberman2005}. In the context of social dynamics, different species can represent different conventions, and a person will imitate the conventions of others after interacting with them. Here, {fitness-based imitation} is the key novelty with respect to simpler models (e.g., the Bass model, where imitation is simply due to an interaction), and it is able to encapsulate some aspects of the complex contagion process.

A seminal implementation of evolutionary dynamics is the model of the dissemination of culture, proposed in~\citet{Axelrod1997}. In this model, each individual is characterized by a multi-dimensional state variable ${\bf x_i}(t)=[x_i^1(t),\dots,x_i^m(t)]$, where each of the $m$ entries represents a cultural trait, which can be assumed to be binary $x_i^k(t)\in\{0,1\}$, for $k=1,2,\hdots, m$. Each trait represents an alternative convention on related topics (e.g., different spelling conventions for a word). At each time step, an individual $i\in\mathcal V$ is picked uniformly at random and this individual interacts with one of their neighbors $j\in\mathcal N_i$, selected at random.

\begin{figure}[t]
    \centering
\includegraphics[width=\linewidth]{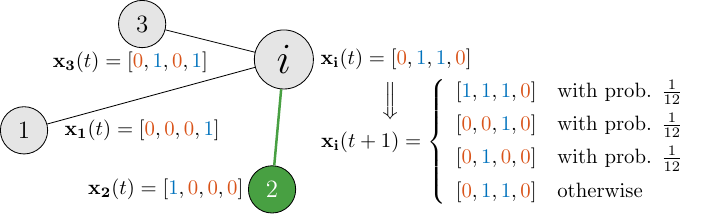}
\caption{Example of an iteration of the evolutionary dynamics in~\citet{Axelrod1997}. 
 Individual $i$ is chosen and has three neighbors: $1$ and $2$ have a single entry of the state in common with $i$ (first and last, respectively), while $3$ has two entries in common (first and second). Assuming that neighbor $2$ is chosen at random (in green), then, $i$ imitates $2$ by copying one of the three entries in which they differ (selected at random), with probability equal to the fraction of entries that they have in common (i.e., $\frac14$), ultimately yielding the update rule written on the right of the figure. }
    \label{fig:evolutionary}
\end{figure}

When $i$ interacts with $j$, $i$ may imitate $j$, depending on how much the two individuals are similar. In particular, with probability proportional to the number of entries of the state on which the two individuals agree, $i$ updates a random entry $\ell$ of their state ${\bf x_i}(t)$ on which $i$ and $j$ disagree (if any), by copying the corresponding entry of $j$, i.e.,  $x_i^\ell(t+1)=x_j^\ell(t)$.  In other words, the more two individuals are culturally similar, the more likely one imitates the other after an interaction on a trait where there is no similarity. All other cultural traits remain unchanged. An explanatory iteration of this update rule is illustrated in Fig.~\ref{fig:evolutionary}.

In its original formulation in~\citet{Axelrod1997}, the model was proposed and studied on regular lattices. Interestingly, numerical simulations for the original model often yielded the emergence of local consensus on the same convention, but global disagreement. In fact,  multiple stable regions characterized by different conventions between different regions were often observed, capturing an important feature of social norms and conventions. Then, building on this seminal work, many efforts have been made to extend the use of evolutionary dynamics in the context of social norms and convention. This body of literature includes establishing analytical results for some specific implementations~\citep{Lanchier2015,Pedraza2021}, generalizing the model and its results to complex networks~\citep{Klemm2003,Guerra2010}, proposing different implementations, e.g., tailored to the emergence of innovation in linguistic conventions~\citep{Baronchelli2006}, and designing control strategies to promote diffusion of innovation over status quo~\citep{Zino2023}.

\subsubsection{Game-theoretic models}\label{sssec:game_models}

The models described above capture some features of complex contagion, but often overlook an important aspect: in social interactions, individuals act strategically in a rational manner. That is, the adoption or otherwise of a certain innovation is the result of a decision-making process, through which an individual tends to achieve some notion of a payoff or reward. This is the underlying rationale beyond the development of mathematical models grounded in evolutionary {game theory}. 

In a game-theoretic framework, each individual is seen as a player, who has two possible actions ($0$ and $1$), corresponding to the state of the individual $x_i$, as in \eqref{eq:state}. Each player $i\in\mathcal V$ who has an interaction with a player $j$ receives a payoff that depends on the state of individual $i$ ($x_i$) and the state of individual $j$ ($x_j$). This payoff
can be encoded in a {payoff matrix}:
\begin{equation}
    \label{eq:payoff_matrix}
     A=\begin{bmatrix}  a\quad &b\\c\quad &d \end{bmatrix},
\end{equation}
 with $a,b,c,d\in\mathbb R$ being constants. Namely, player $i$ receives a payoff equal to $a$ (or $b$) for selecting action $x_i=0$ against an opponent $j$ who plays $x_j=0$ (or $x_j=1$); and a payoff equal to $c$ (or $d$) for selecting action $x_i=1$ against an opponent who plays $x_j=0$ (or $x_j=1$)~\citep{vonNeumann}. For network games, the total payoff that individual $i$ receives for selecting action $0$ (denoted by $u_i(0, {\bf x})$) or $1$ (denoted by $u_i(1, {\bf x})$) is simply given by the average of all payoffs that $i$ would receive from each of their neighbors~\citep{Jackson2015}, i.e.,
\begin{subequations}\label{eq:payoff_function}
\begin{align}
         u_i(0,{\bf x})&=\frac{1}{d_i}\Big(a|\{j\in\mathcal N_i:x_j=0\}|+b|\{j\in\mathcal N_i:x_j=1\}|\Big)=\frac{1}{d_i}\sum_{j\in\mathcal N_i}\Big(a(1-x_j)+bx_j\Big),\\ 
         u_i(1,{\bf x})&=\frac{1}{d_i}\Big(c|\{j\in\mathcal N_i:x_j=0\}|+d|\{j\in\mathcal N_i:x_j=1\}|\Big)=\frac{1}{d_i}\sum_{j\in\mathcal N_i}\Big(c(1-x_j)+dx_j\Big).
\end{align}
 \end{subequations}
Note that the definition $u_i(\cdot,{\bf x})$ highlights that the payoff is dependent on the action (state) of others in the network. Clearly, this approach can be extended, e.g., by associating weights to each edge of the network and transforming \eqref{eq:payoff_function} into a weighted sum, so that each peer 
 provides a different contribution to the payoff of $i$.

\begin{figure}[t]
    \centering
\includegraphics[width=\linewidth]{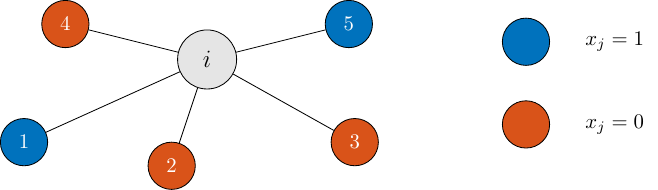}  \caption{Example of a network coordination game. Individual $i$ has five neighbors: three of them are currently adopting the status quo (in orange), and two the innovation (blue). Hence, $u_i(0,{\bf x})=3$ and $u_i(1,{\bf x})=2+2\alpha$. Clearly, $i$ would get a larger payoff for adopting the innovation if and only if $\alpha>\frac12$. }
    \label{fig:coordination}
\end{figure}

 In the context of social norms and convention, {coordination games} are often used to capture the human tendency to conform. In fact, in the most basic implementation of coordination games, the payoff matrix $A$ is the diagonal matrix
\begin{equation}
    \label{eq:coordination}
     A=\begin{bmatrix}  1\quad&0\\0\quad&1+\alpha \end{bmatrix},
\end{equation}
where the parameter $\alpha\in(-1,\infty)$ captures the relative advantage (if positive) or disadvantage (if negative) of the innovation with respect to the status quo. In other words, an individual receives a unit reward for coordinating with a peer on the status quo, and a reward $1+\alpha$ for coordinating on the innovation, as illustrated in Fig.~\ref{fig:coordination}. For the coordination game with payoff matrix in \eqref{eq:coordination}, the payoff functions in \eqref{eq:payoff_function} reduce to the following expressions:
\begin{subequations}\label{eq:payoff_coordination}
\begin{align}
         u_i(0,{\bf x})&=\frac{1}{d_i}\sum_{j\in\mathcal N_i}(1-x_j),\\ 
         u_i(1,{\bf x})&=\frac{1}{d_i}(1+\alpha)\sum_{j\in\mathcal N_i}x_j.
\end{align}
 \end{subequations}

Individuals revise their action (i.e., decide whether to adopt the status quo, $0$, or the innovation, $1$) with the aim of maximizing their payoff. In its simplest implementation, one can assume that individuals are fully rational and always decide to choose the action that gives the maximal payoff, given the action of the others, i.e., to adopt a {best-response} update rule:
\begin{equation}\label{eq:best_response}
 x_i(t+1)=\text{argmax}_{s\in\{0,1\}}u_i(s,{\bf x}(t)).   
\end{equation}
In the case of a tie-break situation, one can presuppose a tie-breaker rule, e.g. to stick with the current action $x_i(t)$. The update rule in \eqref{eq:best_response}, ultimately induces a threshold-like dynamics as in \eqref{eq:ltm}, where an individual's threshold is determined by the parameters in the payoff matrix. For the coordination game in \eqref{eq:coordination}, notice that from \eqref{eq:best_response}, it follows that $x_i(t+1)=1$ if and only if $\text{argmax}_{s\in\{0,1\}}u_i(s,{\bf x}(t))=1$, i.e., if $u_i(1,{\bf x})> u_i(0,{\bf x})$. By substituting the expressions from \eqref{eq:payoff_coordination} into this condition, one obtains that $u_i(1,{\bf x})> u_i(0,{\bf x})$ if and only if \begin{equation}
    \frac{1}{d_i}\sum_{j\in\mathcal N_i}x_j>\frac{1}{2+\alpha}.\
\end{equation}
Evidently, this yields \eqref{eq:ltm} with threshold  $\theta_i=\frac{1}{2+\alpha}$, as in~\citet{morris2000contagion}, shaped by the relative advantage  parameter $\alpha$ of the coordination game. However, note that in the linear threshold model, \eqref{eq:ltm} does not permit agent~$i$ to switch from $x_i(t) = 1$ to $x_i(t+1) = 0$, unlike \eqref{eq:best_response}. Thus, the linear threshold model is more suited for some scenarios where one cannot revert to the status quo, such as installing solar panels on a home, whereas the game-theoretic model is  relevant for norms where one can switch back and forth, e.g. the spelling of a word.

For the sake of completeness, it is worth noticing that the payoff matrix of the coordination game in \eqref{eq:coordination} is quite general. In particular, for the best-response update rule in \eqref{eq:best_response}, the payoff matrix in  \eqref{eq:coordination} is representative of all payoff matrices with $a>c$ and $d>b$. In fact, from \eqref{eq:best_response}, one observes that $x_i(t+1)=1$ if and only if $u_i(1,{\bf x})>u_i(0,{\bf x})$. Using the general expression of the payoff function in \eqref{eq:payoff_function}, one obtains
\begin{equation}
        \frac{1}{d_i}\sum_{j\in\mathcal N_i}\Big(c(1-x_j)+dx_j\Big)>\frac{1}{d_i}\sum_{j\in\mathcal N_i}\Big(a(1-x_j)+bx_j\Big),
 \end{equation}
which can be re-written as
\begin{equation}
        c-a+(a-b-c+d)\frac{1}{d_i}\sum_{j\in\mathcal N_i}x_j>0. 
 \end{equation}
 This in turn reduces to the following condition:
 \begin{equation}
     \frac{1}{d_i}\sum_{j\in\mathcal N_i}x_j>\frac{a-c}{a-c+d-b}=\frac{1}{1+\frac{d-b}{a-c}}.
 \end{equation}
 Hence, any game with payoff matrix as given in \eqref{eq:payoff_matrix}, with $a>c$ and $d>b$, can be reduced to an equivalent coordination game of the form in \eqref{eq:payoff_coordination}, by setting the relative advantage equal to $\alpha=\frac{d-b}{a-c}-1$. For more details on this derivation, one can refer to \citet{zino2022nexus}.

\begin{figure}[t]
    \centering
    \subfloat[small $\beta_i$]{\includegraphics[height=6cm]{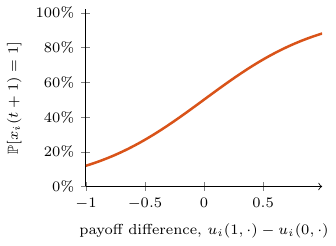}}\quad    
    \subfloat[large $\beta_i$]{\includegraphics[height=6cm]{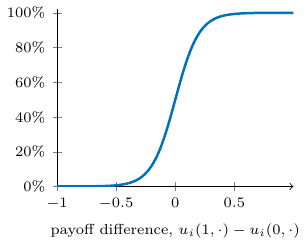}}
 \caption{Probability of adopting the innovation $x_i(t+1)=1$ according to the log-linear dynamics in \eqref{eq:logit} as function of the payoff difference $u_i(1,\cdot)-u_i(0,\cdot)$, for (a) small and (b) large values of the rationality parameter $\beta_i$. }
    \label{fig:logit}
\end{figure}

More complex dynamics have been proposed for the state update, which allows to account for bounded rationality~\citep{mas2016}. In particular, substantial attention has been devoted to the study of a noisy version of the best-response update rule, where it is assumed that individuals revise their state in a probabilistic fashion, according to a {log-linear learning} rule~\citep{blume1995}, also known as the Gibbs model, i.e., 
\begin{equation}\label{eq:logit}
 \mathbb P[x_i(t+1)=s]=\frac{\exp\{\beta_iu_i(s,{\bf x}(t))\}}{\exp\{\beta_iu_i(1,{\bf x}(t))\}+\exp\{\beta_iu_i(0,{\bf x}(t))\}}, 
\end{equation}
where $\beta_i\geq 0$ is a non-negative parameter that captures the rationality of individual~$i$. In particular, when $\beta_i=0$, individual~$i$ will pick either $0$ or $1$ uniformly at random; whereas, in the limit $\beta_i\to\infty$, \eqref{eq:logit} reduces to \eqref{eq:best_response}, and $i$ always selects the action that gives the maximal payoff. For intermediate values of the rationality parameter $\beta_i\in(0,\infty)$, \eqref{eq:logit} posits that individual $i$ adopts the best-response action with a probability greater than $1/2$ but less than $1$, depending on the difference between the payoff for adopting the best response action and the payoff for adopting the other action, as illustrated in Fig.~\ref{fig:logit}. In particular, the larger $\beta_i$ is, the larger the probability that $i$ adopts the action with larger payoff. From another perspective, given a fixed $\beta_i$, the greater the difference between the payoff of the two actions, the greater the probability of selecting the best response action. 

The role of the rationality parameter $\beta_i$, the relative advantage $\alpha$, and the network structure have been extensively investigated for \eqref{eq:logit} in several papers, including~\citet{montanari2010spread_innovation,
young2011dynamics}, where the presence of clusters and cohesive sets has been highlighted as a key factor to favor the adoption of innovation. Alternatively, the diffusion of innovation can be tied to the presence of committed individuals who consistently choose the innovation ($x_i(t)=1$ for all $t\geq 0$), including the number and location in the network~\citep{ye2021nat,Gao2023}; more details are presented below. Similar to Axelrod's model presented above, it is also possible to observe local conformity (local in the sense of the network neighborhood) and global diversity, whereby different clusters in the network adopt $0$ or $1$~\citep{Lambiotte2007}. 

One of the main advantages of this game-theoretic formulation for not only modeling social norms and conventions but social dynamics more generally, is its flexibility. For instance, one can change the structure of the payoff matrix in~\eqref{eq:payoff_matrix} to capture other types of two-player, two-strategy games besides the coordination game, such as anti-coordination and prisoner's dilemma games, to model other types of social dynamics; see~\citep[Appendix~C]{riehl2018survey}. One can also add terms to the function in~\eqref{eq:payoff_function} to account for further behavioral mechanisms besides conformity, such as the role of emerging trends~\citep{zino2022nexus}, existing beliefs or preferences~\citep{peytonyoung2015social_norms}, the presence of anti-conformist individuals~\citep{ramazi2016networks,Vanelli2020}, and social influence~\citep{zino2020chaos,Aghbolagh2023}. Moreover, further sources of complexity can be incorporated into game-theoretic models by considering update rules in which agents have partial information on the game structure, such as mechanisms based on imitation~\citep{Como2021,Govaert2021}, and by extending the set of possible actions to more than two~\citep{DalForno2012}.

\section{Data integration}\label{sec:data}

After having presented some of the main mathematical approaches used to model the formation and evolution of social norms and conventions, this section illustrates how data science has changed the picture around modeling of social dynamics and how different types of data have been integrated towards refining existing approaches and developing novel models validated by data. This section showcases three examples illustrating three different approaches to data-based mathematical modeling. The first example concerns the empirical validation of a threshold model, and a description of how data can be used to inform the thresholds for different types of individuals. The second example considers an evolutionary dynamics model tailored to a linguistic conventions problem, and focuses on the design and conduct of an experiment to validate model-based predictions on tipping points in social conventions. The third example takes a step further in the integration of data in the modeling phase, illustrating how experiments can be designed in order to understand the key terms that should be added to a coordination game in order to faithfully model the formation of a consensus on a convention both at the individual level and at the population level.

\subsection{Data-based determination of thresholds for innovation diffusion}\label{sssec:data_cascade}

\begin{figure}[t]
    \centering
\includegraphics[width=\linewidth]{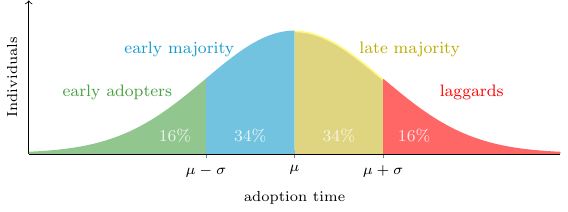}
\caption{Categories of individuals in the classical diffusion innovation theory~\citep{rogers2003diffusion}, where $\mu$ and $\sigma$ are mean and standard deviation, respectively. }
    \label{fig:adoption}
\end{figure}

In his seminal paper~\citet{valente1996social}, Thomas W. Valente used available data on  the diffusion of innovation to validate a threshold-like model, demonstrating how population-level models can fail in capturing important features of the adoption process. To this aim, he built on the categories of early adopters, early majority, late majority, and laggards, extensively used in the theory of diffusion of innovation~\citep{ryan1943diffusion_corn,rogers2003diffusion}. In the classical {innovation diffusion theory} it is assumed that the distribution of the time to adoption across the population is a Gaussian distribution, which ultimately yields the famous S-shaped adoption curve. Under this assumption, individuals involved in the adoption process can be divided into four categories: \begin{enumerate}
    \item early adopters ($\approx16\%$), who adopt the innovation earlier than one standard deviation before the average time of adoption;
    \item early majority ($\approx34\%$), who adopt the innovation after the early adopters, but before the average time of adoption;
        \item late majority ($\approx34\%$), who adopt the innovation after the early majority, but one standard deviation before the average;
    \item laggards ($\approx16\%$), who adopt the innovation at least one standard deviation after the average time of adoption;
\end{enumerate} 
as illustrated in Fig.~\ref{fig:adoption}. For the sake of completeness, note that in some more recent works~\citep{rogers2003diffusion}, a fifth category of ``innovators'' is sometimes used to refer to early adopters who adopt the innovation earlier than one standard deviation before the average time of adoption.

However, these categories refer to the timing at which an individual adopts the innovation with respect to the whole population. In his work, Valente proposed to focus on the proportion of adopters among the neighbors of a selected individual at the time said individual adopts the innovation, which is term {exposure at adoption}, which is nothing but an empirical estimate of the threshold in a linear threshold model. Similar to time of adoption, individuals can be divided into four categories depending on their exposure at adoption, namely, individuals with
\begin{enumerate}
 \item very low threshold ($\approx16\%$), who have exposure at adoption one standard deviation smaller than the average;
    \item  low threshold ($\approx34\%$), who have exposure at adoption greater than those with very low threshold, but smaller than the average;
        \item high threshold ($\approx34\%$), who have exposure at adoption greater than the average but smaller than one standard deviation greater than the average;
    \item very high threshold ($\approx16\%$), who have exposure at adoption one standard deviation greater than the average.\end{enumerate}

In~\citet{valente1996social}, the author considered three datasets involving the adoption of an innovation, for which not only population-level data and individual-level data were available, but also data on social interactions between individuals is present, in order to reconstruct the underlying social network. Specifically, the three datasets concerned i) the prescription of a new drug (tetracycline) by physicians in Illinois in the 1950s, 
 2) the adoption of hybrid corn in Brazilian villages in 1966, and  3) family planning methods for married women in Korea in 1973. From the analysis of all three datasets, Valente found a highly statistically significant correlation between the two categories mentioned above, used to distinguish individuals. That is, early adopters are more likely to have a very low threshold of exposure at adoption, early majority individuals have a low threshold, et cetera. However, on average, only slightly more than half of the individuals were observed to be in the very same category in both classifications. For instance, an individual may belong to the laggard category either because they have a very large threshold, or because they had a low threshold but due to the way in which the innovation spread through the network, they did not receive any exposure until much later. In conclusion, classical categories based on population-level adoption may fail in capturing important aspects of individual-level adoption processes, especially when nontrivial network structures govern the interaction processes.

 The results of this analysis confirm that population models (e.g., the Bass model and its generalizations), despite being able to reproduce the observed S-shaped adoption curves, may neglect important aspects related to how the structure of the social network actually shapes the adoption process. This limits their practical prediction ability, providing strong motivation for the use of agent-based models, which are instead able to describe the process at the individual-level, and thus incorporate the impact of the social network.

\subsection{Experimental determination of tipping points in social conventions}\label{sssec:data_evolutionary}

A problem of paramount importance in the context of social norms and conventions is to understand whether a sufficiently large committed minority can change the societal norm. Committed minority are individuals who promote the innovation actively and consistently when the innovation is still new. Empirical evidence~\citep{Kuran1995}, supported by mathematical models~\citep{Xie2011}, suggested that there is a {critical mass} that determines a {tipping point}: a committed minority smaller than this critical mass cannot overturn an existing status quo, while a committed minority greater than this critical mass is sufficient to trigger social change and the collective adoption of a novel social convention. Several empirical and model-based studies have focused on determining this critical mass, with results spanning from 3.5\% to 40\%~\citep{chenoweth2011civil,Grey2006}. 

In~\citet{centola2018experimental_tipping}, the authors aimed to design an experiment in order to empirically validate the results obtained via mathematical models. Specifically, they considered a problem related to linguistics, where groups of people  have to achieve a consensus on the name of one or multiple objects. For this classical linguistic problem, a mathematical model inspired by evolutionary dynamics has been proposed termed the {Naming Game} model~\citep{Baronchelli2006}. Briefly, each individual~$i$ is characterized by a state $x_i(t)$ that is called the inventory, consisting of word-object pairs. Initially, all inventories are empty. At each time step $t\in \mathbb N_{\geq 0}$, the following procedure is iterated:
\begin{enumerate}
\item[1)] A pair of (neighbors) individuals $i,j$ are picked at random and one
of them plays as speaker (say $i$) and the other as listener;
    \item[2)] The speaker selects randomly an object and retrieves a word from the inventory $x_i(t)$ associated with the object, or, if the inventory is empty, invents a new word;
\item[3a)] If the listener has the word-object pair named by the speaker in the inventory $x_j(t)$, both individuals maintain in their inventories at time $t+1$ only the winning word (i.e, the one word presented by the speaker), deleting all other words associated with the same object;
\item[3b)] If the listener does not have the word presented by the speaker in the inventory, the listener updates the inventory at time $t+1$ by adding the word presented by the speaker.
\item[4)] The time-step is incremented to $t+1$ and the process resumes from item 1).
\end{enumerate}

\citet{centola2018experimental_tipping} proposed the following experimental setting to validate this model. They recruited 194 participants to play an online repeated game, divided into 10 different groups. In the first stage, in each group, participants were randomly coupled and they were shown a picture of a face and are asked to insert the name of that face. If the players entered the same name, they
were given a monetary reward, otherwise they were penalized; this procedure was iterated, until the whole population reached a consensus on the name of that face. Then, in a second stage of the experiment, in each of the 10 groups, a different fraction of computer bots were introduced (from 15\% to 35\%), who acted as a committed minority, supporting a novel name instead of the established status quo. Then, the same procedure of random matches (including also bots) and monetary rewards/punishments were iterated. 

The results of this experiment confirmed the model-based predictions from \citet{Baronchelli2006}. In fact, calibrating the parameters of the model (namely, number of objects, population size, and number of iterations) to the experimental setting, one obtained that the critical mass was approximately 25\%. In \citet{centola2018experimental_tipping}, all groups  with less than 25\% bots showed only a very small minority of participants would eventually switch to the novel name. On the contrary, groups with at least 25\% bots showed 72\% uptake of the new name. Interestingly, the experiments provide empirical support to the use of agent-based models to derive not only qualitative observations concerning formation and evolution of social norms and conventions, but also in determining precise quantitative predictions.

\subsection{Key factors in decision-making via online experiments}\label{sssec:data_games}

In mathematical models built on coordination games, as described above, individuals tend to make decisions on which action to take among two (or more) conventions by maximizing a payoff function that depends solely on how many of their neighbors adopt a certain convention, as per Eqs.~(\ref{eq:payoff_function}--\ref{eq:coordination}). However, while social coordination and the tendency to conform with others are key factors in human decision-making concerning social norms and conventions~\citep{marques1994}, social scientists have observed and theorized the presence of other behavioral factors that play a role in individual-level decision-making. The social and behavioral science community agree on the presence of two important factors, namely the presence of {inertia} (sometimes referred to as status-quo bias) and the sensitivity to {trends}. Concerning the first aspect, individuals often prefer to stick to their current decisions and have some inherent resistance to change even if there are clear benefits to switching~\citep{samuelson1988statusquo}. The second aspect, instead, is related to the fact that people tend to follow the trends observed in the population by being more attracted behaviors or products that are becoming more popular, even if it is currently only adopted by a minority~\citep{sparkman2017dynamicnorm_sustainable}. These two factors are well known in the social psychology literature, and many researchers have empirically validated their presence in individual-level decision making in different contexts. However, they are not present in the mechanism of coordination game models, potentially limiting the ability of such models to faithfully reproduce real-world phenomena.

\begin{figure}[t]
    \centering
\subfloat[]{\includegraphics[height=5cm]{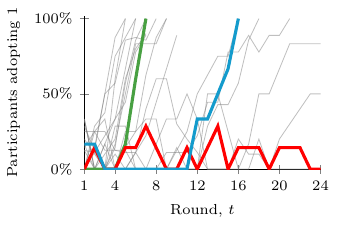}\label{fig:exp1}}\quad\subfloat[]{\includegraphics[height=5cm]{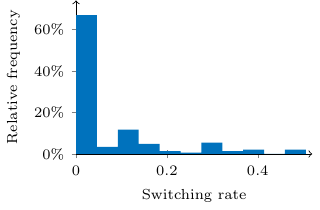}\label{fig:exp2}}
\caption{Results of the experiment to determine key factors in decision-making concerning the adoption of conventions from \citet{ye2021nat}. Panel (a) illustrates different adoption curves for the innovation, illustrating scenarios of fast (green) and delayed (blue) adoption, and no adoption (red). Panel (b) shows the switching rate, which is used in \citet{ye2021nat} to calibrate the model. }
    \label{fig:exp}
\end{figure}

In~\citet{ye2021nat}, the authors built on these social psychology observations and theories to design an {experimental paradigm} to investigate whether inertia and sensitivity to trends have downstream consequences on the emergent behavior of the population concerning the formation and evolution of conventions, and thus to determine whether it is important to incorporate these factors in a game-theoretic model. In the experiments, participants were enrolled online and divided into small groups (each group comprised of $12$ agents: $8$--$10$ human participants and $2$--$4$ computer bots). In each group, participants made repeated decisions, choosing between two options (say, $0$ or $1$), with the final goal of reaching a consensus. At each round, each participant was able to see the proportion of the rest of the group that chose each option in the previous round. Computer bots were added to initially enforce a majority on a status quo (say $0$), and then flipped to try to promote the adoption of the innovation (say $1$) by selecting the other option. The results of the experiment are summarized in Fig.~\ref{fig:exp}. From a statistical analysis of the experimental results conducted based on individual-level data, the authors  concluded that coordination, inertia, and sensitivity to trends were all present in the individual-level decision making mechanism. Specifically, the state of an individual $i$ at time $t+1$ was significantly influenced by i) the state of others at time $t$, ii) their own current state $x_i(t)$, and iii) whether the number of adopters of the alternative had increased or decreased over the previous iteration of the game. For more details on the experimental paradigm, see~\citet{Mlakar2024}.

Building on this experimental evidence, the authors in~\citet{ye2021nat} conjectured that individuals revise their state according to the log-linear learning rule in \eqref{eq:logit}, with payoff functions that extend those of a coordination game in \eqref{eq:payoff_coordination} to account for inertia and sensitivity to trends. In particular, it was proposed that:
\begin{subequations}\label{eq:social_model_payoff}
\begin{align}
u_i(1,{\bf x}(t))&= \frac{b_i}{d_i}\sum_{j\in\mathcal N_i} x_j(t) + k_i x_i(t) +r_i 
\hat x_i(t)\,, \\
u_i(0,{\bf x}(t))&= \frac{b_i}{d_i}\sum_{j\in\mathcal N_i} \big(1-x_j(t)\big) + k_i \big(1-x_i(t)\big) +r_i \big(1-\hat x_i(t)\big)\,,
\end{align}\end{subequations}
where \begin{equation}\label{eq:prediction}\hat x_i(t) =  \frac{1}{2}\Big(1+\frac{1}{n-1}\sum_{j\in\mathcal{V}\setminus\{v\}} \big(x_j(t) - x_j(t-1)\big)\Big)\end{equation} 
captures the trend, being $\hat x_i(t)>\frac12$ if the fraction of adopters of action $1$ has increased in the previous time-step and $\hat x_i(t)<\frac12$ otherwise. The payoff in \eqref{eq:social_model_payoff}  expands \eqref{eq:payoff_coordination} by adding two terms. In fact, the first term captures coordination with no relative advantage for the innovation, and coincides with the standard term from \eqref{eq:payoff_coordination} with $\alpha=0$. The second term ($k_ix_i$ and $k_i(1-x_i)$, respectively) increases the payoff for keeping the current action. The third term ($k_ix_i$ and $k_i(1-x_i)$, respectively)  increases the payoff for selecting the action whose support has increased in the previous time-step. 

The three parameters $b_i, k_i, r_i$ are non-negative scalar constants, weighting the contribution of the three terms, and can be assumed to sastify $b_i+k_i+r_i = 1$ without loss of generality. In~\citet{ye2021nat}, these parameters were calibrated by fitting the individual-level experimental data illustrated in Fig.~\ref{fig:exp2}, assuming that there are two classes of individuals, and with individuals of the same class having the same parameters to avoid overfitting. In other words, by exploring a grid on the parameter space, one could determined the configuration that minimized the discrepancy between the number of times individuals change their action (from $0$ to $1$ or vice versa) in the simulations and the same quantity recorded from experimental data. This calibration demonstrated that, in the absence of the two additional terms (i.e, by enforcing $k_i=0$ and/or $r_i=0$), a model calibrated with individual-level data fails in capturing the range of population-level emergent behavior observed in the experiment (see Fig.~\ref{fig:exp1}). On the contrary, when all three terms are included into \eqref{eq:social_model_payoff}, the model is able to faithfully reproduce the experimental data at both the individual- and population-level.

In summary, this work describes how experiments can be designed in order to improve existing mathematical models, going beyond a simple calibration of a model to actually informing the very mathematical formulation of the model dynamics and, ultimately, validating its improvement with respect to existing models. This procedure allows to adopt existing modeling paradigms and refine them using experimental data to ultimately derive improved models, which can be used to explore the social phenomena of interest beyond the experimental limitations. An example of this can be found, e.g., in \citet{Gao2023}, where the experimentally-validated model from~\citet{ye2021nat} is used to explore the impact of the network structure on the evolution of conventions.

\section{Summary, conclusions, and vision}

This chapter has provided an overview of the mathematical modeling of social systems, applied directly to the formation, persistence, and evolution of social norms and conventions. After reviewing the basic notions and features of norms and conventions, several different frameworks were presented. Then, the integration of data into these models was discussed, using examples from the literature. These data ranged from those obtained in field surveys to controlled psychology experiments. In this last section, the key challenges of data-driven mathematical models of social systems are discussed, and a future vision is proposed based on potential opportunities. While the following commentary is centered around modeling of social norms and conventions, it largely extends to broader problems, such as modeling of opinion and belief formation, dis/misinformation propagation, and collective action. Addressing these challenges will not only require the advancement of mathematical modeling and analysis techniques, but crucially, close collaboration with colleagues in the theoretical and experimental social and behavioral sciences. 

Despite several decades of sustained development involving efforts of the scientific community spanning multiple disciplines, when it comes to direct applications, mathematical modeling of social systems still lags behind other fields such as ecology and epidemiology. In fact, while the COVID-19 pandemic created global disruption and tragically resulted in millions of deaths~\citep{WHO}, it was also a crowning moment for the real-world application of mathematical models of epidemic spreading~\citep{Vespignani2020,Estrada2020}. A range of different models, from agent-based models to population models, were used to predict COVID-19 spread, and crucially, inform government and help plan public policy and medical interventions~\citep{Giordano2020,DellaRossa2020,Zhao2020,casella2020,parino2021,Truszkowska2021}. A similar systematic and widespread adoption of mathematical models in the context of social systems is still far from our daily life, due to the presence of several challenges that need to be addressed. In the following, two challenges that are especially associated with modeling social systems are discussed.

\subsection{Challenges}

One challenge, perhaps unique to modeling of human behavior (as opposed to e.g., modeling of infectious diseases or predator-prey relations in ecology) is that people are by nature highly context-dependent. For instance, the decision-making processes for collective action and protest movements (whether to participate and with what tactics) share some commonalities with norms and conventions (which norm to adopt), such as the fact that peer pressure and internal preferences can play a key role. However, there are also fundamental differences; a person's willingness to protest is closely tied to how strongly they (internally) identify with the protest movement and group, as well as interactions with the authority~\citep{louis2022failure,louis2020volatility}, while what matters most for norms is to coordination with others and conform to the majority~\citep{Lewis2002,Bicchieri2005}. Thus, one may require a different framework and starting point for modeling protest movements~\citep{thomas2024mobilisation} as opposed to norms and conventions. Separately, the fact the people learn through social interactions and adapt over time is well-known~\citep{fay2010interactive}, and thus may also require modeling~\citep{acemoglu2011bayesian}. These considerations can create a difficult balance between a mathematician's natural instinct to develop models and methods that can generalize across multiple application domains and contexts, and the established view in social and behavioral science that people are highly context-dependent. Finding this balance, and being able to identify the level of {generalizability} of various models, will contribute greatly towards the real-world applicability of mathematical models of social systems. It will require concerted efforts in collaboration with other disciplines, and close scrutiny of how various decision-making and behavioral processes are modeled, and to what degree there is alignment with theoretical foundations and empirical evidence from the behavioral sciences.

Another limit to the use of real-world data to inform mathematical models of social dynamics is the limited possibility to access {high-quality data} of individuals' behaviors. In fact, for many other domains of applications of mathematical models (e.g., physical systems or ecological systems), many tools have been developed for obtaining accurate and reliable estimations of the quantities involved in the dynamics, e.g., physical quantities or amount of individuals for each species in a geographical region~\citep{Tredennick2021}. For social dynamics, while it is indeed true that the ubiquitous nature of social media platforms and the development of data science techniques have allowed to process huge amounts of data, high  quality data are often difficult to obtain. In fact, data is often available at the population level (e.g., the adoption curve), but rarely available at the individual-level (e.g., the pattern of adoption in a social network). Moreover, many datasets are owned by private companies, making it difficult for researchers to access. In addition, data collected via experiments are strongly dependent on the specific experimental setting, so are difficult to re-use in other contexts or generalize. Similar issues related to the data quality were present in epidemic modeling before 2020. However, the global call for standardized and high-quality epidemiological data for COVID-19 has enabled giant leaps in  the field of mathematical epidemiology~\citep{Gardner2021}. A key challenge for the future research in mathematical models of social systems is the development of common guidelines to create repositories of high-quality data of social dynamics, which will be used to improve and inform mathematical models. 

\subsection{Opportunities}

So far, there has been much focus on using experimental data to inform, refine, and validate models. One obvious opportunity is to explore the converse, viz. using calibrated mathematical models to explore and design different {interventions}, which can then be tested using real-world experiments. Identifying ways to create ``social tipping points'' that lead to mass change in norms and conventions has become of increasing interest, seen as a way to address pressing societal issues ranging from the climate crisis to pandemic response~\citep{otto2020social,Coleman2021,Mlakar2024}. There exist many possible levers for interventions, and it is impractical and costly to empirically explore all of these (e.g. with experiments or field tests) to identify the most effective and relevant intervention, especially since interventions can be context-dependent (see above for a separate discussion on this issue). Moreover, many interventions at the individual level are based on social and cognitive psychology methods, and the effect size can be small even if statistically significant; it is difficult to use experiments to establish whether these small effects at the individual level can lead to population level tipping points. Having validated and calibrated mathematical models would allow one to systematically and rapidly (at least relative to planning and running a field experiment) examine an array of interventions, including those that vary in time~\citep{Walsh2023,zino2023_adaptive}. In the modeling framework, one can identify interventions most effective at generating tipping points, including an ``optimal'' intervention by associating each intervention with a cost. These insights could then inform which interventions to test in an empirical setting, and how. Employing mathematical models in this manner is slowly becoming accepted in the social and behavioral sciences community, and such an approach is a major opportunity for mathematicians to contribute in the future. 

Another opportunity for future developments in the mathematical modeling of social conventions and norms lies in the explosive growth of {machine learning} techniques and artificial intelligence. In particular, developments in machine learning have paved the way for the use of mathematical paradigms such as that of universal differential equations~\citep{Rubel1981}, which augment model-based approaches with machine-learnable structures within the general context of {physics-informed machine learning}~\citep{Karniadakis2021}. Briefly, universal differential equations consists of parameterizable systems of equations that do not dictate the precise evolution of a dynamical system, but rather determine certain conditions that any possible evolution must fulfill (e.g., enforcing the emergence of a S-shaped curve in social diffusion). The key aspect is that these  systems of equations can be augmented and trained with deep learning techniques, as proposed in~\citep{Rackauckas2020,Koch2023}. Recently, some research groups have started using these machine learning techniques in the context of developing models of human behavior with applications in epidemiology~\citep{Kuwahara2024}. These new methods may pave the way for a promising direction of future exploration, toward a  systematic methodology that merges theoretically-informed and data-driven approaches to develop effective mathematical models for social dynamics. 

Finally, it is worth noticing that social systems are increasingly integrated within complex cyber-physical systems~\citep{Annaswamy2021}. Hence, also the formation, persistence, and evolution of social norms and conventions are phenomena that are deeply intertwined with physical and technical dynamics. For instance, a paradigm shift towards increasing the use of public transportation is needed to reduce the carbon footprint of our cities. However, human decisions concerning the use of public transportation are clearly dependent on the efficiency of the traffic network, which in turn is affected by the number and type of vehicles present in the system and, in turn, human decisions, thus completing the feedback loop~\citep{Hull2008}. In a different application field, the virality of new trends, memes and products on a social media platform are increasingly guided by recommender systems, which are algorithms that filter the information and content an individual is presented with~\citep{Jannach2010}. In turn, the current decisions of individuals to engage with the content (watching, liking, sharing) is considered by the recommender system in terms of future content to present to individuals. {Cyber-physical-human systems} have emerged as a novel and promising paradigm that integrates a model of human behavior with a model of a cyber-physical system~\citep{Annaswamy2021}, offering a unified approach to represent and study the complex socio-technical systems that are becoming pervasive in our daily life. Given their recent emergence, the exploration of data-based modeling approaches for human behavior within the context of cyber-physical-human systems, building on the techniques discussed in this chapter, is still a research direction mostly unexplored and highly promising.

\end{document}